\documentclass[english,showpacs,floatfix,12pt]{revtex4}
\usepackage[dvips]{graphicx}
\usepackage{longtable}
\usepackage{amsmath,amssymb}
\usepackage{dcolumn}
\usepackage{latexsym}
\usepackage{babel}
\usepackage[latin1]{inputenc}
\usepackage{color}
\usepackage[colorlinks]{hyperref}
\def\pa{\partial}
\def\al{\alpha}
\def\ga{\gamma}

\def\la{\lambda}
\def\La{\Lambda}
\def\be{\beta}
\def\kp{\kappa}
\def\th{\theta}
\def\sg{\sigma}
\def\vf{\varphi}

\def\lan{\langle}
\def\ran{\rangle}
\def\eps{\varepsilon}

\def\l{\left}
\def\r{\right}
\def\nn{\nonumber}
\def\ck{\check}

\def\E{{\cal{E}}}
\def\N{{\cal{N}}}

\begin{document}
\title{Mass Spectrum of Dirac Equation with \\ Local Parabolic Potential}
\author{Ying-Qiu Gu}
\email{yqgu@fudan.edu.cn}  \affiliation{School of Mathematical
Science, Fudan University, Shanghai 200433, China}
\date{6th January 2018}

\begin{abstract}
In this paper, we solve the eigen solutions and mass strectra of the
Dirac equation with local parabolic potential which is approximately
equal to the short distance potential generated by spinor itself.
The mass spectrum is quite different from that of a spinor in
Coulomb potential. The masses of some baryons are similar to this
one. The mass-angular momentum relation $m=m(J,n)$ is quite similar
to the Regge trajectories. The parabolic potential has property of
asymptotic freedom near the center and confinement at large
distance. So the results imply that, the local parabolic potential
may be more suitable for describing nuclear potential approximately.
The solving procedure can also be used to solve the Dirac equation
with other complicated potential.

\end{abstract}
\maketitle

\section{Introduction}
\setcounter{equation}{0}

For hadrons, the relation between mass $m$ and quantum numbers
$(J,n)$ is usually described by the Regge-Chew-Frautschi
formula\cite{1,2},
\begin{eqnarray}
m^2=an+bJ+m_0, \label{rcf}
\end{eqnarray}
where $(a,b,m_0)$ are constants for the exited states of the same
kind particle. In many cases, the factors satisfies $b\le a\le
2b$\cite{3,4}. The Regge trajectory is an important tool widely used
to analyze the spectroscopy of mesons and baryons. Various
theoretical models have been constructed to explain the mass spectra
of particles and to derive the Regge trajectories, such as
non-relativistic quark models \cite{5,6,7,8,9,10,25,[260],11}, flux
tube model or similar string
model\cite{3,12,13,14,15,16,17,18,19,20}, semi-relativistic model
\cite{21,22,23,24}, relativistic  model \cite{26,27,28,29,30},
quantum chromodynamics (QCD) sum rule \cite{31,32,33}, color
hyperfine interaction \cite{34,35,36,37}, Lattice QCD
model\cite{latt,38,39,40,41} and so on. There are also some Regge
phenomenology investigations\cite{3,42,43,44,45,46,47}. By
statistical and regressive method to get the relation $m=f(J,n)$.

In many models, the total potential between quarks is given by
Cornell potential with some hyperfine terms of correction, and the
mass spectrum is solved in relative Jacobi
coordinates\cite{8,10,11,29,30,34,48}. In \cite{48}, by semi
classical approximation and Bohr-Sommerfeld quantization, the
Regge-like relation $E\sim L^{2\al/(\al+2)}$ and $E\sim
n^{2\al/(\al+2)}$ for large $(n,L)$ is derived for power-law
confining potentials $V\propto r^\al$. By the phenomenological
researches, we also find that the Regge-Chew-Frautschi formula
(\ref{rcf}) is only approximately valid, and a little nonlinearity
always exists\cite{42,43,45}. The specific Regge trajectories depend
on concrete confining potential. However, no matter what confining
potential is, the analytic relation $m=f(J,n)$ for the excited
states always exists.

Recently, a number of experimental data for highly exited resonances
were reported \cite{49,50,51,52,53,54,55,56,57}. These data provide
opportunity to check the previous calculations and develop more
effective models. As pointed out in  \cite{57}, A better
understanding of the nucleon as a bound state of quarks and gluons
as well as the spectrum and internal structure of excited baryons
remains a fundamental challenge and goal in hadronic physics. In
particular, the mapping of the nucleon excitations provides access
to strong interactions in the domain of quark confinement. While the
peculiar phenomenon of confinement is experimentally well
established and believed to be true, it remains analytically
unproven and the connection to quantum chromodynamics (QCD) - the
fundamental theory of the strong interactions - is only poorly
understood. In the early years of the 20th century, the study of the
hydrogen spectrum has established without question that the
understanding of the structure of a bound state and of its
excitation spectrum need to be addressed simultaneously. The
spectroscopy of excited baryon resonances and the study of their
properties is thus complementary to understanding the structure of
the nucleon in deep inelastic scattering experiments that provide
access to the properties of its constituents in the ground state.

The quark models employ multiplets of spinors and nonlinear
interactive vectors with gauge symmetries, which are too complicated
to get exact solutions and an overview for the properties. In this
paper we examine the following simple and closed Dirac equation with
short range self-generating vector potential $\Phi_\mu$,
\begin{eqnarray}
{\cal L} = \phi^+\al^\mu (i\pa_\mu-s\Phi _{\mu})\phi-\mu c\ck\ga
+\frac 1 2 (\pa_\mu\Phi_\al \pa^\mu\Phi^\al-b^2 \Phi_\mu\Phi^\mu),
\label{lag}
\end{eqnarray}
in which $\ck\ga= \phi^+\ga\phi$. (\ref{lag}) has plentiful spectra.
By the Regge trajectories  we find the excited states may be
relevant to some of baryons.

\section{Equations and  Simplification}
\setcounter{equation}{0}

At first, we introduce some notations. Denote the Minkowski metric
by $\eta_{\mu\nu}={\rm diag}(1,-1,-1,-1)$, Pauli matrices by
\begin{eqnarray}
 {\vec\sg}=(\sg^{j})= \l \{\l (\begin{array}{cc}
 0 & 1 \\ 1 & 0 \end{array} \r),\l (\begin{array}{cc}
 0 & -i \\ i & 0 \end{array} \r),\l (\begin{array}{cc}
 1 & 0 \\ 0 & -1 \end{array} \r)
\frac{}{} \r\}.\label{1.1}\end{eqnarray} Define $4\times4$ Hermitian
matrices as follows
\begin{eqnarray}\al^\mu=\l\{\l ( \begin{array}{cc} I & 0 \\
0 & I \end{array} \r),\l (\begin{array}{ll} 0 & \vec\sg \\
\vec\sg & 0 \end{array}
\r)\r\},\quad \ga =\l ( \begin{array}{cc} I & 0 \\
0 & -I \end{array} \r),\quad \be =\l ( \begin{array}{cc} 0 & -iI \\
iI & 0 \end{array} \r) \label{1.2}
\end{eqnarray}
where $\mu\in\{0,1,2,3\}$, $x^0=ct$ and $\al^\mu=\ga^0\ga^\mu$. By
variation of (\ref{lag}) we get the Dirac equation and dynamics of
$\Phi^\mu$,
\begin{eqnarray}
\al^\mu(\hbar i\pa_\mu-s \Phi_\mu)\phi &=&\mu c\ga\phi.
\label{1.04}\\
\pa_\al\pa^\al \Phi^\mu+b^2\Phi^\mu &=&- s\ck\al^\mu,\quad
\ck\al^\mu= \phi^+\al^\mu\phi.\label{ephi}
\end{eqnarray}
For the eigen states of $\phi$, only the magnetic quantum number
$m_z$ and the sipn $s$ are conserved. So the eigen solution takes
the following form
\begin{eqnarray}
\phi=(u_1,u_2e^{\vf i},-iv_1,-iv_2e^{\vf i})^T \exp(m_z \vf
i-\frac{mc^2} \hbar it),  \label{fms}
\end{eqnarray}
where the index $T$ stands for transpose, $ m_z \in\{0,\pm 1,\pm
2,\cdots\}$, and $u_k,v_k(k=1,2)$ are real functions of $r$ and
$\th$. However, the exact solution of (\ref{fms}) does not exist,
and we have to solve it by effective algorithm\cite{gu1,58}. Since
the numerical solutions are also unhelpful to understand the global
structure of the mass spectrum, we seek for the approximate analytic
solutions in this paper.

Different from the case of an electron, a proton has a hard core
with charge distribution, and the radius of the distribution is
about $1\times 10^{-15} m$. The following calculation shows the
local parabolic potential is approximately equal to $\Phi_\mu$ near
the center, then we have
\begin{eqnarray}
\Phi_0\dot =\l(\frac {w^2 r^2}{2\rho^2}-{2(1-\eta)}\r)\mu c,\quad
\vec \Phi \dot = 0, \quad (r<12\rho),
  \label{ptn} \end{eqnarray}
in which $w$ is the strength factor, $\eta$ is a parameter to adjust
the depth of confinement to fit the true confining potential.
$\rho=\frac \hbar {\mu c}$ is the theoretical Compton wave length,
which is used for nondimensionalization of the Dirac equation.

In order to simplify (\ref{lag}), we make transformation\cite{58}
\begin{eqnarray} g=u_1+u_2 i\qquad f=v_1-v_2 i.
\label{gf}
\end{eqnarray}
Substituting (\ref{fms}), (\ref{ptn}) and (\ref{gf}) into
(\ref{lag}) we get Lagrangian as
\begin{eqnarray}
{\cal L}  =  ({\cal L}_0 + {\cal L}_f)\mu c, \label{lagt}
\end{eqnarray}
in which we defined
\begin{eqnarray}
 {\cal L}_0 &\equiv & \Re \lan e^{\th i}\left( -\bar g
(\pa_r+\frac i r \pa_\th) f + f (\pa_r+\frac i r \pa_\th) \bar g
\right)\ran\rho
-\frac i {r\sin\th}(m_z+\frac 1 2 ) (\bar g\bar f-gf)\rho \nn \\
&&+\eps (|g|^2+|f|^2) +\l(\frac {w^2 r^2}{2\rho^2}-2(1-\eta)\r)|g|^2
-(2+\kp)|f|^2, \label{lag0}
\end{eqnarray} where $\Re\lan\ran$ stands for taking real part, and
\begin{eqnarray}
{\cal L}_f \equiv \l( \kp +\frac {w^2 r^2}{2\rho^2}-{2(1-\eta)}
\r)|f|^2. \label{lage}
\end{eqnarray} In (\ref{lag0}) $\eps$ is relative mass defect defined by
\begin{eqnarray}
m c^2 =(1-\eps)\mu c^2,\qquad \eps =\frac {\mu-m}{\mu}, \label{epd}
\end{eqnarray}
and  $\rho$ is used as length unit, $\kp$ is a constant to let
$|\int_0^\infty{\cal L}_f r^2 dr| \to 0$ so that convergent rate of
the procedure is optimized. In the case (\ref{ptn}) we set
$\kp=(1-\eta)$ which is about the mean value of the potential in the
effective domain, and then ${\cal L}_f$ can be omitted for the $0th$
order approximation. For proton we have
\begin{eqnarray}
\rho=\frac \hbar {\mu c}=(1-\eps)\frac \hbar{m_p c},~~\la =\frac
\hbar{m_p c}= 2.1037\times 10^{-16}{\rm m}. \label{lam}
\end{eqnarray}

In (\ref{lagt}), ${\cal L}_0$ almost keeps all invariance of
relativity and has simple and complete eigensolutions, which can be
used as the bases of Hilbert space of representation. ${\cal L}_f$
is the trouble terms with small energy, which acts as perturbation
in the calculation. If taking $\mu c=1, \rho=1$, (\ref{lagt})
becomes dimensionless.

For (\ref{lag0}), the rigorous eigensolutions take the following
form\cite{58}
\begin{eqnarray}
g=U(r) [P(\th)+Q(\th)i],\qquad f=V(r) [P(\th)+Q(\th)i] e^{-i\th}.
\label{gfuv}\end{eqnarray} By variation of (\ref{lag0})  we get
\begin{eqnarray}
\pa_\th P &=& \cot\th m_z P+(m_z+K)Q,\label{eqp}\\
\pa_\th Q &=& -\cot\th(m_z+1)Q+(m_z+1-K)P ,
\label{eqq}\end{eqnarray} in which $K=\pm1,\pm2,\cdots$
corresponding to orbital angular momentum, $P,Q$ are associated
Legendre functions. The radial functions satisfy
\begin{eqnarray}
\pa^2_r U +\frac 2 r \pa_r U - \l(\frac {K(K-1)}{r^2}-\frac {2{\cal
E}}{\rho^2}+\frac {\l(3-\eps-\eta\r)w^2r^2}{2\rho^4}\r)U=0 ,
\label{equ}\end{eqnarray}
\begin{eqnarray} V=\frac{\l(r\pa_r
U-(K-1)U\r)\rho}{(3-\eps-\eta)r}, \label{eqv}\end{eqnarray} in which
we defined
\begin{eqnarray}
{\cal
E}&=& \frac 1 2(2-\eps-2\eta)(3-\eps-\eta)\label{epse}\\
&=& \frac 1 2 \l((1-2\eta)(2-\eta)+3(1-\eta) M+ M^2\r),
\label{epse1}\\
{\rm Or ~inversely,}~~~\eps&=&\frac1 2
\l(5-3\eta-\sqrt{(1+\eta)^2+8{\cal E}}\r),
\end{eqnarray} where $M=\frac m \mu$ is dimensionless mass. The
above equations can be easily solved, and the solutions are all
elementary functions. The normalizing conditions are as follows
\begin{eqnarray}
\int_0^\pi (P^2+Q^2) 2\pi \sin\th d\th=1,\quad \int_0^\infty
(U^2+V^2) r^2 d r=1.  \label{norm}\end{eqnarray}

\section{Eigen Solutions to the equation}
\begin{figure}[ht]
\centering
\includegraphics[width=12cm]{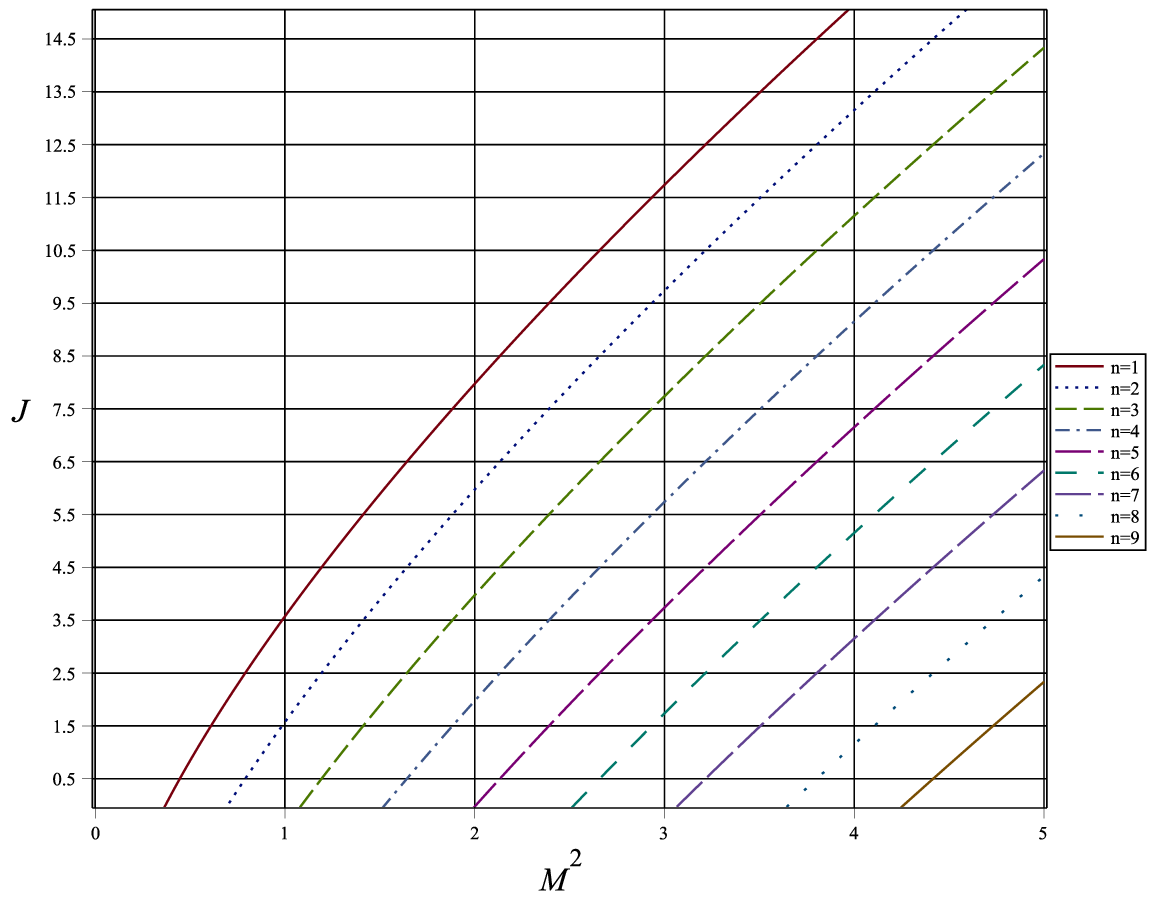}
\caption{Regge trajectroies of (\ref{regg}). Each intersection
between lines $J=\frac 1 2 +j$ and curves $J=J(n,M^2)$ corresponds
to one or more particles, and we have about 90 intersections in the
figure. Considering  degenerate states, the figure contains more
than 1000 particles with different quantum numbers $(n, K, m_z,
s)$.} \label{fig0}
\end{figure}

\setcounter{equation}{0} For (\ref{equ}), we have the solution
\begin{eqnarray}
U&=&\l( C_1  r^{K-1}  + C_2 r^{-K} \r)L_{n-1}^{J}\l(\frac {2r^2}
{r_n^2}\r) \exp\l(-\frac {r^2} {r_n^2}\r),
\label{slu} \\
n&=&1,2,3,\cdots,\qquad  J=\l|K-\frac 1 2\r|=\frac 1 2, \frac 3
2,\frac 5 2,\cdots,
\end{eqnarray} where $L_{n-1}^J$ is associated Laguerre polynomials,
$n$ is radial quantum number, and $J$ is angular momentum quantum
number. $C_1=0$ corresponding to $K<0$ and $C_2=0$ corresponding to
$K>0$. The energy spectrum and radius parameter is given by
\begin{eqnarray}
N w &=& \frac {{\cal E}} {\sqrt{1+\eta+\sqrt{(1+\eta)^2+8{\cal
E}}}},\qquad N= n +\frac 1 2({J-1}),\label{eps} \\
r_n &=&  2\rho \sqrt{\frac {N}{{\cal E}}}= {2M\la} \sqrt{\frac
{N}{{\cal E}}}. \label{rds}\end{eqnarray}  Substituting
(\ref{epse1}) into (\ref{eps}) we get Regge-like relation as follows
\begin{eqnarray}
2n +J-1 = \frac {\sqrt2} { 2 w}\l(1-2\eta+M\r)\sqrt{2-\eta+M}.
\label{regg}\end{eqnarray} Or inversely,
\begin{eqnarray}
M&=&\frac 1{36} {\N}^{\frac 2 3}+4{ { \left( 1+\eta \right)
^{2} }{{\N}^{-\frac 2 3}}}+\frac 1 3 (5\eta-4),\label{mnj}\\
\N &=&216\,Nw\sqrt {2}+24\,\sqrt {162\,{N}^{2}{w}^{2}-3\, \left(
1+\eta \right) ^{3}}.\label{mnj1}
\end{eqnarray}
In (\ref{regg}) we have 3 constants $(w,\eta,\mu)$ for the same
series of particles to be determined by empirical data. Although the
form of (\ref{regg}) or (\ref{mnj}) is quite different from
(\ref{rcf}), the following calculation shows that the curves of
(\ref{regg}) in the effective domain is quite near straight
lines(see Fig.\ref{fig0}).

Substituting (\ref{slu}) and (\ref{eps}) into (\ref{eqv}), we can
derive $V$. By calculation we get
\begin{eqnarray}
\int_0^\infty U^2_{K,n} r^2 d
r=\frac{(1+\eta)\sqrt{(1+\eta)^2+8{\cal E}}+(1+\eta)^2+4{\cal E}}
{(1+\eta)\sqrt{(1+\eta)^2+8{\cal E}}+(1+\eta)^2+6{\cal E}}.
\label{uvn}\end{eqnarray} For all meaningful eigensolutions we have
$0.1<\E<4$, and then we have $\int_0^\infty U^2_{K,n} r^2 d
r=0.8\sim 1$. Therefore, the relative truncation error for the $0th$
approximation is about $10\%$.

The ground state corresponds to $n=1,J=\frac 1 2$, and then we have
$N=\frac 3 4$. Considering energy degeneracy, we only need to
calculate the energy spectrums while $K\ge 1$. For the ground state
of proton, we have empirical data $r_n \sim 1\times 10^{-15}{\rm m}$
and $m_p=938.28$MeV. Substituting them and $N=\frac 3 4$ into
(\ref{epd}), (\ref{lam}),(\ref{epse}) and (\ref{rds}), we get
constants $(w,\eta,\rho,\mu,\eps_0)$ expressed by ${\cal E}_0$. If
taking ${\cal E}_0 = 0.17618$, we have solution
\begin{eqnarray}
\eta=0.74238,~w=0.11972,~\eps_0=0.33220,~\rho = 1.4048\times
10^{-16}{\rm m},~\mu c^2 = 1.4051{\rm GeV}.
\label{dat}\end{eqnarray} For a proton, by (\ref{dat}) we find $r_n
\sim 7\rho$. By (\ref{epd}) and $\eps_0$ we find the relative mass
defect of strong interaction confinement is about $33\%$. The
observational mass $m_p$ is much less than constant mass $\mu$. This
case is quite different from an electron without strong interaction.

Substituting (\ref{dat}) and $(J,n)$ into (\ref{regg}), (\ref{mnj})
and (\ref{mnj1}), we get the mass spectra of the particles $m(J, n)$
shown in Tab.\ref{table-1}. We find the masses of many baryons are
near the spectra, and (\ref{regg}) is quite similar to the Regge
trajectories of baryons(see Fig.\ref{fig0}). By (\ref{eps}) and
(\ref{rds}), we find the radius parameter $r_n$ of the excited
states even decreases a little when the quantum number $N$
increases. Detailed calculation shows we always have $\bar r=(4\sim
10)\rho$ for all particles. This means a particle with local
parabolic potential or short distance potential $\Phi_\al$ has a
very hard core. This phenomenon is quite different from the case of
Coulomb potential, where we have $\bar r\propto n^2$.
\begin{table}
\caption{Mass spectra of Dirac equation with local parabolic
potential(MeV)} \label{table-1}
\begin{tabular}{lllllllllllllllllllll}
\hline
$m(J,n)$  &   1   &   2   &   3   &   4   &   5   &   6   &   7   &   8   &   9   &   10  &   11  &   12  &   13  &   14  \\
\hline
1/2 &   938 &   1249    &   1535    &   1801    &   2052    &   2291    &   2520    &   2740    &   2953    &   3159    &   3359    &   3554    &   3744    &   3930    \\
1+1/2   &   1098    &   1395    &   1670    &   1928    &   2173    &   2407    &   2631    &   2847    &   3056    &   3260    &   3457    &   3650    &   3838    &   4022    \\
2+1/2   &   1249    &   1535    &   1801    &   2052    &   2291    &   2520    &   2740    &   2953    &   3159    &   3359    &   3554    &   3744    &   3930    &   4112    \\
3+1/2   &   1395    &   1670    &   1928    &   2173    &   2407    &   2631    &   2847    &   3056    &   3260    &   3457    &   3650    &   3838    &   4022    &   4202    \\
4+1/2   &   1535    &   1801    &   2052    &   2291    &   2520    &   2740    &   2953    &   3159    &   3359    &   3554    &   3744    &   3930    &   4112    &   4290    \\
5+1/2   &   1670    &   1928    &   2173    &   2407    &   2631    &   2847    &   3056    &   3260    &   3457    &   3650    &   3838    &   4022    &   4202    &   4378    \\
\hline
6+1/2   &   1801    &   2052    &   2291    &   2520    &   2740    &   2953    &   3159    &   3359    &   3554    &   3744    &   3930    &   4112    &   4290    &   4465    \\
7+1/2   &   1928    &   2173    &   2407    &   2631    &   2847    &   3056    &   3260    &   3457    &   3650    &   3838    &   4022    &   4202    &   4378    &   4552    \\
8+1/2   &   2052    &   2291    &   2520    &   2740    &   2953    &   3159    &   3359    &   3554    &   3744    &   3930    &   4112    &   4290    &   4465    &   4637    \\
9+1/2   &   2173    &   2407    &   2631    &   2847    &   3056    &   3260    &   3457    &   3650    &   3838    &   4022    &   4202    &   4378    &   4552    &   4722    \\
10+1/2  &   2291    &   2520    &   2740    &   2953    &   3159    &   3359    &   3554    &   3744    &   3930    &   4112    &   4290    &   4465    &   4637    &   4806    \\
\hline
11+1/2  &   2407    &   2631    &   2847    &   3056    &   3260    &   3457    &   3650    &   3838    &   4022    &   4202    &   4378    &   4552    &   4722    &   4889    \\
12+1/2  &   2520    &   2740    &   2953    &   3159    &   3359    &   3554    &   3744    &   3930    &   4112    &   4290    &   4465    &   4637    &   4806    &   4972    \\
13+1/2  &   2631    &   2847    &   3056    &   3260    &   3457    &   3650    &   3838    &   4022    &   4202    &   4378    &   4552    &   4722    &   4889    &   5054    \\
14+1/2  &   2740    &   2953    &   3159    &   3359    &   3554    &   3744    &   3930    &   4112    &   4290    &   4465    &   4637    &   4806    &   4972    &   5135    \\
15+1/2  &   2847    &   3056    &   3260    &   3457    &   3650    &   3838    &   4022    &   4202    &   4378    &   4552    &   4722    &   4889    &   5054    &   5216    \\
\hline
\end{tabular}
\end{table}

We find the masses of many baryons are near the spectra. Obviously
each excited state should correspond to an observable particle. This
means some baryons can be regarded as excited resonances of a
proton.  How to exactly identify the quantum numbers for each
particle observed in experiments is an important but fallible
problem.

As the $0th$ order approximation with only 3 free coefficients, the
result is satisfactory. To get more accurate solutions of
(\ref{lag}), we can expand $\phi$ as series of the eigen functions
of (\ref{lag0}) and then solve mass spectra of (\ref{lag})\cite{58}.
However, in this case we have only numerical results without an
overview on the spectra.

\section{Effectiveness of the Parabolic Potential}
\setcounter{equation}{0}

Now we check the effectiveness of the parabolic potential for
nuclear potential. It is well known the global parabolic potential
cannot be used as confining potential of Dirac equation. However,
the following calculations show the local parabolic potential is
effective to describe nuclear potential approximately.

\begin{figure}[ht]
\centering
\includegraphics[width=12cm]{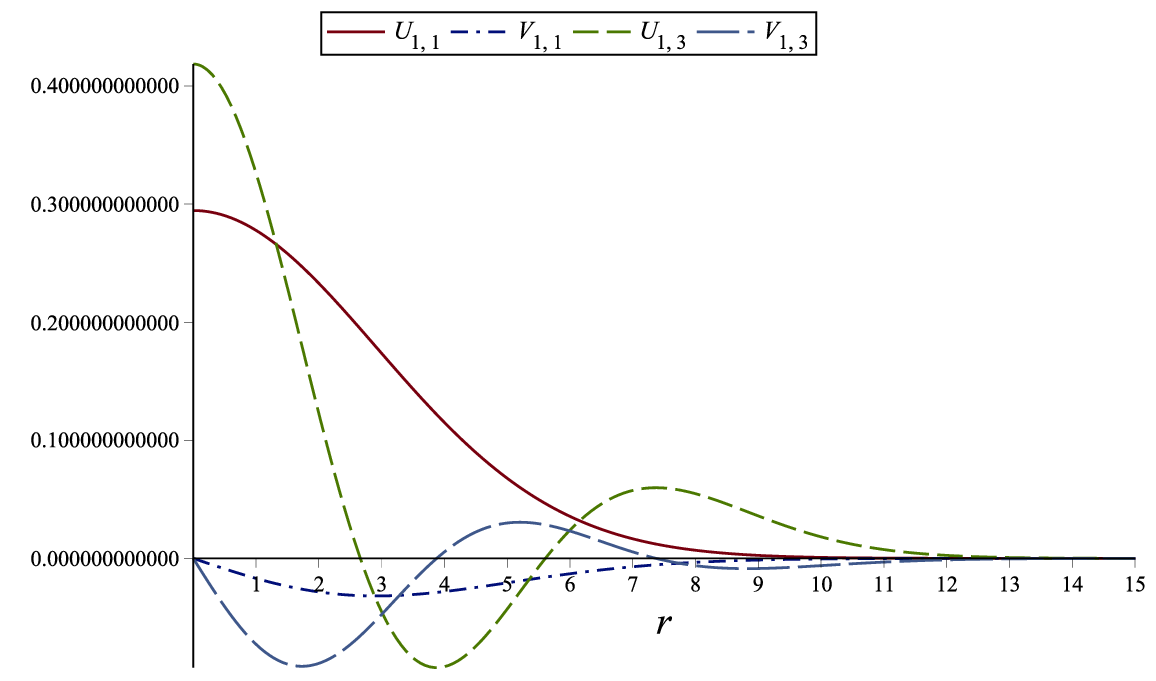}
\caption{Some radial wave functions of a spinor in local parabolic
potential.} \label{fig1}
\end{figure}

\begin{figure}[ht]
\centering
\includegraphics[width=12cm]{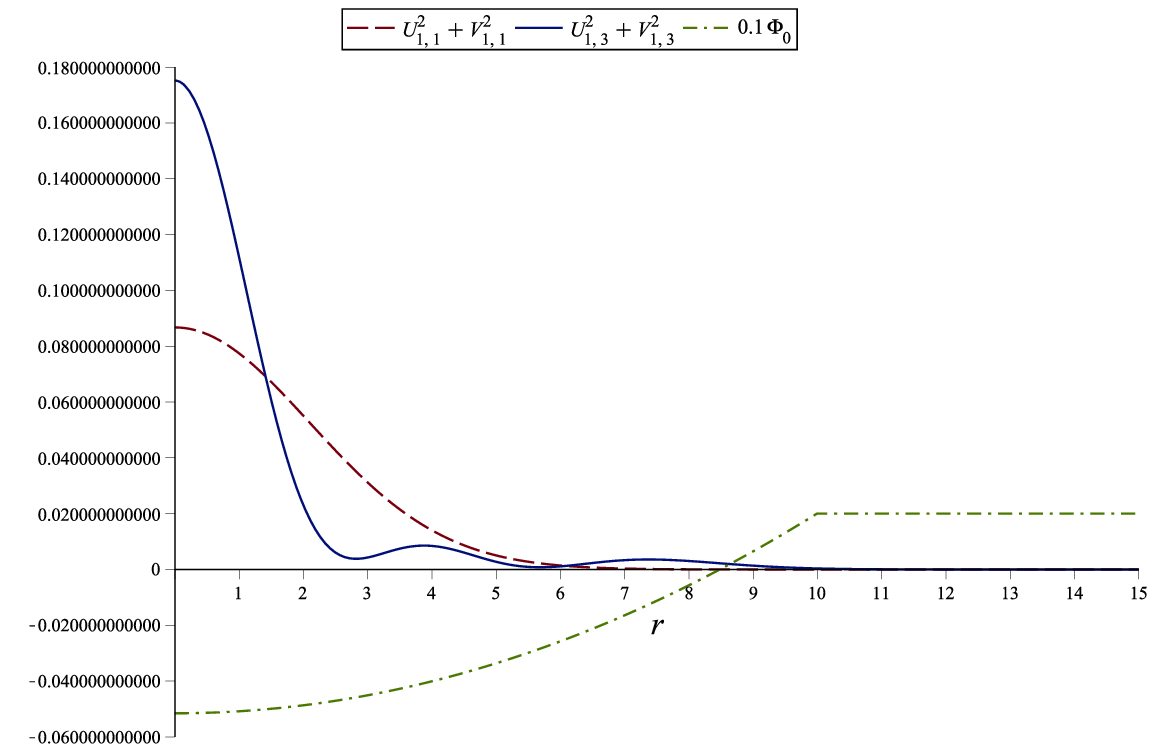}
\caption{ The effective domain of local parabolic potential and
modules of radial wave functions. The spinor with short distance
potential is mainly concentrated near the center, and it does not
diffuse when $K$ or $n$ increases.} \label{fig2}
\end{figure}

At first, we check all radial functions $(U_{K,n},V_{K,n})$  and
their module $U^2+V^2$ are almost distributed in the domain $r<12$,
where is also the effective area of the local parabolic potential.
The first couple of the radial functions is given by
\begin{eqnarray}
U_{1, 1} = 0.29450687e^{-0.05873333r^2},\quad V_{1, 1} =
-0.01796750r e^{-0.05873333r^2}, \label{uvf}\end{eqnarray} we find
$|V/U|^2\sim 1\%$. See Fig.\ref{fig1} and Fig.\ref{fig2} as follows,
where we take $\rho=1$ as length unit.

\begin{figure}[ht]
\centering
\includegraphics[width=12cm]{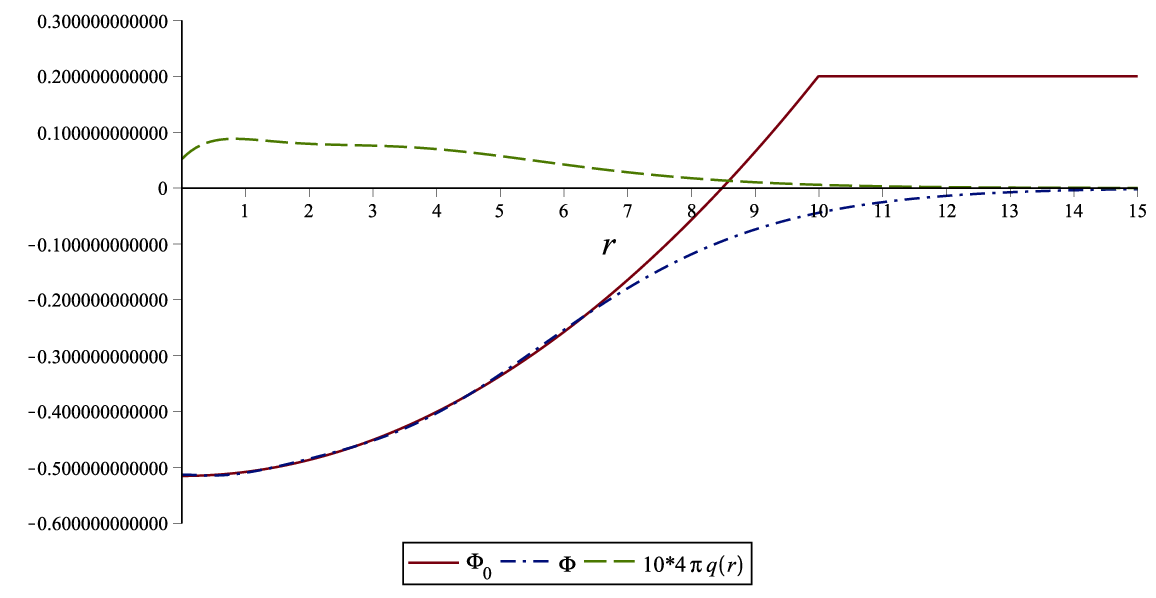}
\caption{The parabolic potential versus the short range potential
generated by $q(r)$.} \label{fig3}
\end{figure}

Secondly, for the following short range potential $\Phi$ with source
$q(r)$,
\begin{eqnarray}
\pa_\al\pa^\al \Phi+\frac 1 {\rho^2}\Phi=  -65\cdot\frac{ 4 \pi
q(r)}{\rho^3}. \label{ephi0}\end{eqnarray} By Fig.\ref{fig3}, we
find the solution $\Phi$ is almost parabolic potential $\Phi_0$ in
the domain $r<8$ for the above source, for which the normalizing
condition is $4\pi \int q(r)r^2 dr=1$. This means in the interior of
a baryon, the potential of strong interaction may be different from
the Cornell potential or  potential generated by point source
$-\frac {e^{-r}}{r}$ or the MIT bag model. It may be more suitably
described by local parabolic potential.

Thirdly, by (\ref{eps}) we find that, different from electron in
Coulomb potential, in this case the radius parameter $r_n$ of the
wave function even decreases a little as the increasing of quantum
number $N$. So the local parabolic potential is also suitable for
the excited states.

\section{Discussion and Conclusion}
\setcounter{equation}{0}

As the $0th$ order approximation, the above calculation provides
some important messages. The Dirac equation with short distance
potential has quite different energy spectrum and eigen functions
from that with Coulomb potential. Dirac equation is a magic equation
with marvellous properties which should be strictly
analyzed\cite{eng,gu2,engc}. The nuclear potential may be more
similar to local parabolic potential, rather than the MIT bag model
or $-\frac {e^{-r}} r$ or $-\frac {\al_s} r - ar$. Obviously, the
spinor in short distance potential (\ref{ephi0}) has the asymptotic
freedom near the center and strong confinement  near $r\sim
10\rho$(see Fig.\ref{fig2} and Fig.\ref{fig3}).

To get more accurate results, we should directly solve the coupling
system of (\ref{ephi}) and (\ref{1.04}), and expand the radial
functions $(U,V)$ upon the bases $U_{K,n},V_{K,n}$ of the
representation space\cite{58}. However, in this case we have not
explicit analytic expression (\ref{eps}) for mass spectra.

As the alternative models for fundamental particles, some simple and
closed systems such as the following one are worth to be carefully
studied,
\begin{eqnarray}
{\cal L}&=&\phi^+\al^\mu (i\pa_\mu-eA_\mu-s\Phi _{\mu})\phi-\mu
c\ck\ga
 +F(\ck\ga,\ck\be)\nn\\
&~& -\frac 1 2 \pa_\mu A_\al\pa^\mu A^\al +\frac 1 2
(\pa_\mu\Phi_\al \pa^\mu\Phi^\al-b^2 \Phi_\mu\Phi^\mu).
\label{prtn1}
\end{eqnarray}
Some deep secrets may be concealed under the nonlinear potential $F$
and short distance potentials, because the spinor equation is a
magic equation.

If we denote $\phi\dot =\psi_{-1}+\psi_0+\psi_1$, where $\psi_k$ are
basis eigenfunctions in the Hilbert space of representation, and
$\psi_0$ is the main component. Substituting them into (\ref{prtn1})
and using the orthogonality of $\psi_k$, we get
\begin{eqnarray}
{\cal L}&\dot=&\sum_{k=-1}^1\l(\psi_k^+\al^\mu
(i\pa_\mu-eA_\mu-s\Phi
_{\mu})\psi_k-\mu c\ck\ga_k+F(\ck\ga_k,\ck\be_k) \r)\nn\\
&~& -\frac 1 2 \pa_\mu A_\al\pa^\mu A^\al +\frac 1 2
(\pa_\mu\Phi_\al \pa^\mu\Phi^\al-b^2 \Phi_\mu\Phi^\mu)+ G(\psi_k,
A^\al, \Phi^\al). \label{quak}
\end{eqnarray}
In (\ref{quak}) $(\psi_{-1},\psi_0,\psi_1)$ may be easily
interpreted as quarks with fraction electric charge and confinement,
and the cross terms $G$  may be interpreted as gauge fields. For any
complicated mathematical models a little vigilance should be
remained, because Nature only uses simple but best mathematics, and
the complicated equations easily lead to inconsistence and
singularity.

On the other hand, the  regression analysis  for empirical data to
derive mass function with single integer variable $m=m(N),~
(N=1,2,3,\cdots)$ for similar particles is much important, because
like Hydrogen spectra $E_N=\frac 1 2 \hbar \omega \al^2 N^{-2}$,
such analytic function certainly exists and is usually very simple,
and then to determine the further relation between quantum numbers
$N=a n+b J +m_0$ is relatively easy. This procedure need not to
concern the physical meanings of $N$ at first and gets rid of the
fallible and misleading task to identify the quantum numbers $n$ and
$J$ for each particles at the beginning. If we can arrange the
masses of similar particles from small to large at each horizontal
integer coordinates $N=1,2,3,\cdots$ to get smooth curves, the
regressive function $m=m(N)$ for all smooth curves can be derived.
From the final mass function $m=m(a n+b J +m_0)$ of high precision,
we can determine the specific potentials in Dirac equation
conversely.

\end{document}